\documentclass[11pt]{article}
\usepackage{color,graphicx,amssymb,amsfonts}
\usepackage[mathscr]{eucal}

\newcommand{\SetFigFont}[3]{}

\title{A Lattice Model for the Fermionic Projector \\
in a Static and Isotropic Space-Time}
\author{Felix Finster and W\"atzold Plaum}
\date{December 2007}

\newtheorem{Def}{Definition}[section]

\newtheorem{Prp}[Def]{Proposition}

\newcommand{\Proof}{{\em{Proof.}}}
\newcommand{\QED}{\ \hfill $\FBox$ \\[1em]}

\newcommand{\spc}{\;\;\;\;\;\;\;\;\;\;}

\newcommand{\bra}{\mbox{$< \!\!$ \nolinebreak}}
\newcommand{\ket}{\mbox{\nolinebreak $>$}}

\newcommand{\R}{\mathbb{R}}
\newcommand{\1}{\mbox{\rm 1 \hspace{-1.05 em} 1}}
\newcommand{\Z}{\mathbb{Z}}

\newcommand{\slsh}{\mbox{ \hspace{-1.1 em} $/$}}
\newcommand{\vslsh}{\mbox{$v$ \hspace{-1.15 em} $/$}}
\newcommand{\Tr}{\mbox{\rm{Tr}\/}}
\newcommand{\tr}{\mbox{\rm{tr}\/}}

\newcommand{\beq}{\begin{equation}}
\newcommand{\eeq}{\end{equation}}
\newcommand{\FBox}{\rule{2mm}{2.25mm}}

\newcommand{\floc}{f_{\mbox{\scriptsize{loc}}}}

\newcommand{\Lat}{{\mathfrak{L}}}

\begin{document}
\maketitle

\begin{abstract}
We introduce a lattice model for a static and isotropic
system of relativistic fer\-mions.
An action principle is formulated, which describes a
particle-particle interaction of all fermions.
The model is designed specifically for a numerical analysis
of the nonlinear interaction, which is expected to lead to the
formation of a Dirac sea structure.
We discuss basic properties of the system.
It is proved that the minimum of the variational principle is attained.
First numerical results reveal an effect of spontaneous symmetry breaking.
\end{abstract}

\section{Introduction} \label{sec1}
It is generally believed that the concept of a space-time continuum
(like Minkowski space or a Lorentzian manifold) should be modified
for distances as small as the Planck length. The principle of the
fermionic projector~\cite{PFP} proposes a mathematical framework for
physics on the Planck scale in which space-time is discrete. The
physical equations are formulated via a variational
principle for fermionic wave functions defined on a finite
set of space-time points, without referring to notions like space,
time or causality. The idea is that these additional structures,
which are of course essential for the description of nature, arise
as a consequence of the nonlinear interaction of the fermions as
described by the variational principle. More specifically, it was
proved that the original permutation symmetry of the space-time
points is spontaneously broken by the fermionic wave
functions~\cite{F3}. This means that the fermions will induce
non-trivial relations between the space-time points. In particular,
one can introduce the notion of a ``discrete causal structure'' (see
the short review article~\cite{F5}). The conjecture is that for
systems involving many space-time points and many particles, the
fermions will group to a ``discrete Dirac sea structure'', which in
a suitable limit where the number of particles and space-time points
tends to infinity, should go over to the well-known Dirac sea
structure in the continuum. Then the ``discrete
causal structure'' will also go over to the usual causal structure
of Minkowski space~\cite{PFP}.

Hints that the above conjecture is true have been obtained coming
from the continuum theory. First, our variational principle has a
well defined continuum limit~\cite[Chapter~4]{PFP}, and we get
promising results for the resulting effective continuum
theory~\cite[Chapters~6-8]{PFP}. Furthermore, rewriting certain
composite expressions ad hoc as distributions in the continuum, one
finds that Dirac sea configurations can be stable minima of our
variational principle~\cite[Chapter~5.5]{PFP}. The ad-hoc procedure
of working with distributions is justified in the paper~\cite{F4},
which also gives concrete hints on how the regularized fermionic
projector should look like on the Planck scale. For a more detailed
stability analysis in the continuum see~\cite{FH}.

Despite these results, many questions on the relation between discrete space-time
and the continuum theory remain open. In particular, it seems an important task to
complement the picture coming from the discrete side; that is, one should analyze
large discrete systems and compare the results with the continuum analysis.
Since minimizing the action for a discrete system can be regarded as a problem of non-linear
optimization, numerical analysis seems a promising method.
Numerical investigations have been carried out successfully for small systems involving few particles
and space-time points~\cite{FSD}. However, for large systems, the
increasing numerical complexity would make it necessary to use more sophisticated
numerical methods or to work with more powerful computers.
Therefore, it seems a good idea to begin with simplified systems,
which capture essential properties of the original system but are easier to
handle numerically.
In this paper, we shall introduce such a simplified system. The method is
to employ a spherically symmetric and static ansatz for the fermionic projector.
This reduces the number of degrees of freedom so much that
it becomes within reach to simulate systems which are so large that they
can be compared in a reasonable way to the continuum.

The paper is organized as follows. In Section~\ref{sec2} we review the
mathematical framework of the fermionic projector in discrete space-time
and introduce our variational principle. In Section~\ref{sec3} we
take a spherically symmetric and static ansatz in Minkowski space and discretize
in the time and the radial variable to obtain a two-dimensional lattice.
In Section~\ref{sec4} our variational principle is adapted to this two-dimensional setting.
In Section~\ref{sec5}
we give a precise definition of our model and discuss its basic properties;
for clarity this section is self-contained and independent of the rest of the paper.
In Section~\ref{sec6} the existence of minimizers is proved.
In Section~\ref{sec7} we present first numerical results and discuss
an effect of a spontaneous symmetry breaking.
We point out that the purpose of this paper is to define the model and
to discuss some basic properties. Numerical simulations of larger
systems will be presented in a forthcoming publication.

\section{A Variational Principle in Discrete Space-Time} \label{sec2}
We briefly recall the mathematical setting of discrete space-time and
the definition of our variational principle in
the particular case of relevance here (for a more general introduction see~\cite{F1}).
Let~$H$ be a finite-dimensional complex vector space endowed with a non-degenerate
symmetric sesquilinear form~$\bra .|. \ket$. We call~$(H, \bra .|. \ket)$ an
{\em{indefinite inner product space}}.
The adjoint~$A^*$ of a linear operator~$A$ on~$H$ can be defined
as in Hilbert spaces by the relation~$\bra A \Psi \,|\, \Phi \ket
=\bra \Psi \,|\, A^* \Phi \ket$. A selfadjoint and idempotent
operator is called a {\em{projector}}.
To every element~$x$ of a finite set~$M=\{1, \ldots, m\}$ we associate
a projector~$E_x$. We assume that these projectors are orthogonal and
complete,
\beq \label{complete}
E_x\,E_y \;=\; \delta_{xy}\:E_x\:,\spc \sum_{x \in M} E_x \;=\; \1\:.
\eeq
Furthermore, we assume that the images~$E_x(H) \subset H$ of these
projectors are all four-dimensional and non-degenerate of
signature~$(2,2)$.
The points~$x \in M$ are called {\em{discrete space-time points}}, and the corresponding
projectors~$E_x$ are the {\em{space-time projectors}}. The
structure~$(H, \bra .|. \ket, (E_x)_{x \in M})$ is
called {\em{discrete space-time}}. Furthermore, we introduce
the {\em{fermionic projector}}~$P$ as a projector on a
subspace of~$H$ which is negative definite and of dimension~$f$.
The vectors in the image of~$P$ have the interpretation as the
occupied quantum states of the system,
and~$f$ is the number of particles. We refer
to~$(H, \bra .|. \ket, (E_x)_{x \in M}, P)$ as a {\em{fermion
system in discrete space-time}}.

When forming composite expressions in the projectors~$P$ and~$(E_x)_{x \in M}$,
it is convenient to use the short notations
\beq \label{notation}
\Psi(x) \;=\; E_x\,\Psi \spc {\mbox{and}} \spc P(x,y) \;=\; E_x\,P\,E_y \:.
\eeq
Using~(\ref{complete}), we obtain for any~$\Psi, \Phi \in H$ the formula
\beq \label{local}
\bra \Psi \:|\: \Phi \ket \;=\; \sum_{x \in M} \bra \Psi(x)\:|\: \Phi(x) \ket_{E_x(H)} \:,
\eeq
and thus the vector~$\Psi(x) \in E_x(H) \subset H$ can be thought of as the
``localization'' of the vector~$\Psi$ at the space-time point~$x$.
Furthermore, the operator~$P(x,y)$ maps~$E_y(H) \subset H$ to~$E_x(H)$, and it is often
useful to regard it as a mapping only between these subspaces,
\[ P(x,y)\;:\; E_y(H) \: \rightarrow\: E_x(H)\:. \]
Again using~(\ref{complete}), we can write the vector~$P \Psi$ as follows,
\[ (P\Psi)(x) \;=\; E_x\: P \Psi \;=\; \sum_{y \in M} E_x\,P\,E_y\:\Psi
\;=\; \sum_{y \in M} (E_x\,P\,E_y)\:(E_y\,\Psi) \:, \]
and thus
\beq \label{diskernel}
(P\Psi)(x) \;=\; \sum_{y \in M} P(x,y)\: \Psi(y)\:.
\eeq
This relation resembles the representation of an operator with an integral kernel.
Therefore, we call~$P(x,y)$ the {\em{discrete kernel}} of the fermionic projector.

To introduce our variational principle, we define the
{\em{closed chain}}~$A_{xy}$ by
\beq \label{cdef}
A_{xy} \;=\; P(x,y)\: P(y,x) \;:\; E_x(H) \rightarrow E_x(H)\:.
\eeq
Let~$\lambda_1,\ldots,\lambda_4$ be the zeros of the characteristic polynomial
of~$A_{xy}$, counted with multiplicities. We define the {\em{Lagrangian}} by
\beq \label{Ldef}
{\mathcal{L}}[A_{xy}] \;=\; \frac{1}{8} \sum_{i,j=1}^4 \left(
|\lambda_i|-|\lambda_j| \right)^2
\eeq
and introduce the {\em{action}} by summing over the space-time points,
\beq \label{Sdef}
{\mathcal{S}}[P] \;=\; \sum_{x,y \in M} {\mathcal{L}}[A_{xy}]\:.
\eeq
Our variational principle is to minimize this action under
variations of the fermionic projector. We remark that
this is the so-called critical case of the
auxiliary variational principle as introduced in~\cite{PFP, F1}.

\section{The Spherically Symmetric Discretization} \label{sec3}
Recall that in discrete space-time, the subspace~$E_x(H) \subset H$ associated to a space-time
point~$x \in M$ has signature~$(2,2)$. In the continuum, this vector space is to
be identified with an inner product space of the same signature: the space of
Dirac spinors at a space-time point~$x \in \R^4$ with the inner product~$\overline{\Psi} \Phi$,
where~$\overline{\Psi}=\Psi^\dagger \gamma^0$ denotes the adjoint spinor.
For any $4 \times 4$-matrix~$B$ acting on the spinors,
the adjoint with respect to this inner product is denoted by~$B^*=\gamma^0 B^\dagger \gamma^0$. Furthermore, the indefinite inner product
space~$(H, \bra .|. \ket)$ in the continuum should correspond to the space
of Dirac wave functions in space-time with the inner product
\beq \label{ip}
\bra \Psi \:|\: \Phi \ket \;=\; \int \overline{\Psi(x)}\, \Phi(x)\: d^4x\:.
\eeq
This resembles~(\ref{local}), only the sum has become a space-time integral integral.
Likewise, in~(\ref{diskernel}) the sum should be replaced by an integral,
\[ (P \Psi)(x) \;=\; \int P(x,y)\, \Psi(y)\: d^4y\:, \]
where now~$P(x,y)$ is the integral kernel of the fermionic projector of the continuum~$P$.
Since we assume that our system is isotropic, it follows that it is
{\em{homogeneous}} in space. Furthermore, we assume that our system is {\em{static}}, and thus
the integral kernel depends only on the difference~$y-x$,
\beq \label{homogeneous}
P(x,y) \;=\; P(\xi) \spc {\mbox{for all~$x,y \in \R^4$ and~$\xi := y-x$}}.
\eeq
We take the Fourier transform in~$\xi$,
\beq \label{fourier}
P(\xi) \;=\; \int \frac{d^4k}{(2 \pi)^4}\: \hat{P}(p)\: e^{i p \xi}\:,
\eeq
where~$p \xi$ denotes the Minkowsi inner product of signature~$(+--\,-)$.
Let us collect some properties of~$\hat{P}(p)$. First, the operator~$P$
should be symmetric (= formally self-adjoint) with respect to the
inner product~(\ref{ip}). This means for its integral kernel that
\beq \label{tsymm}
P(\xi)^* = P(-\xi) \:,
\eeq
and likewise for its Fourier transform that
\[ \hat{P}(p)^* \;=\; \hat{P}(p)\:. \]
Assuming as in~\cite[\S4.1]{PFP} that the fermionic projector has a
{\em{vector-scalar structure}}, ${\hat{P}}$ can be written as
\beq \label{vss}
\hat{P}(p) \;=\; \hat{v}_j(p) \,\gamma^j + \hat{\phi}(p)\,\1
\eeq
with a real vector field~$\hat{v}$ and a real scalar field~$\hat{\phi}$.
Moreover, the assumption of {\em{spherical symmetry}} implies
that the above functions depend only on~$\omega := p^0$ and on~$k:=|\vec{p}|$,
and that the vector component can be written as
\[ \hat{v}_j \gamma^j \;=\; \hat{v}_0 \,\gamma^0 + \hat{v}_k \, \gamma^k
\qquad {\mbox{with}} \qquad \gamma^k \;:=\; \frac{\vec{p} \,\vec{\gamma}}{|\vec{p}|} \]
and real-valued functions~$\hat{v}_0$ and~$\hat{v}_k$.
Next we can exploit that the image of~$P$ should be
{\em{negative definite}}. Moreover, since~$P$ should be a projector,
it should have {\em{positive spectrum}}. Since in Fourier space, $P$
is simply a multiplication operator, we can consider
the operator~$\hat{P}(p)$ for any fixed~$p$. This gives rise to the conditions that the vector
field~$\hat{v}$ must have the same Lorentz length as~$\hat{\phi}$ and must be
past-directed,
\[ \hat{v}_0\;<\; 0 \qquad {\mbox{and}} \qquad \hat{v}_0^2 - \hat{v}_k^2 \;=\; \hat{\phi}^2 \:, \]
and furthermore that~$\hat{\phi}$ must be non-negative.
Combining the above conditions, we conclude that~$\hat{P}$ can be written in
the form
\beq \label{momentum}
\hat{P}(p) \;=\; \hat{\phi}(\omega, k) \,\Big( \1 \,-\, \gamma^0 \,\cosh \tau(\omega, k)
\,+\, \gamma^k \,\sinh \tau(\omega, k) \Big)
\eeq
with a non-negative function~$\hat{\phi}$ and a real function~$\tau$.
Note that we have not yet used that~$P$ should be idempotent, nor that
the rank of~$P$ should be equal to the number of particles~$f$. Indeed, implementing
these conditions requires a more detailed discussion, which we postpone
until the end of this section.

We next compute the Fourier transform of~(\ref{momentum}), very similar as in~\cite[Lemma~5.1]{F4}.
Introducing in position space the polar coordinates~$\xi = (t,r,\vartheta, \varphi)$
and assuming that~$r \neq 0$, the scalar component becomes
\begin{eqnarray*}
\phi(t,r)&=& \frac{1}{(2 \pi)^4}\int_{- \infty}^{\infty}d\omega\,\int_0^{\infty} k^2\, dk
\int_{-1}^{1} d \cos \vartheta \, \int_{0}^{2 \pi} d \varphi \; \hat{\phi}(\omega,k)\,e^{i \omega t - i k r \cos \vartheta} \\
&=& \frac{1}{4 \pi^3\, r} \int_{-\infty}^{\infty}\, d \omega\, e^{i \omega t}
\int_{0}^{\infty} k\, dk\: \sin(kr)\;\hat{\phi}(\omega,k)\:.
\end{eqnarray*}
The zero component of the vector component is computed similarly,
\[ v_0(t,r) \;=\; -\frac{\gamma^0}{4 \pi^3\, r} \int_{-\infty}^{\infty}\, d \omega\, e^{i \omega t}
\int_{0}^{\infty}\,k\, dk\: \sin(kr)\;\hat{\phi}(\omega,k)\: \cosh \tau(\omega,k) \:. \]
For the calculation of the radial component, we first need to pull the Dirac matrices out of the integrals,
\begin{eqnarray*}
\lefteqn{ \int \frac{\mbox{d}^4p}{(2 \pi)^4} \:\hat{v}_k \gamma^k
\, e^{i p \xi} \;=\; \frac{\vec{\gamma} \vec{\nabla}_{\vec{x}}}{(2 \pi)^4}
\int \mbox{d}^4p\; \hat{v}_k\left(\omega,k\right)\,\frac{i}{k}\: e^{i \omega t - i \vec{k} \vec{x}} } \\
&=& i \vec{\gamma} \vec{\nabla} \left( \frac{1}{4 \pi^3\, r}\:
\int_{-\infty}^{\infty}\, d \omega\, e^{i \omega t}
\int_{0}^{\infty} dk\: \sin(kr)\; \hat{\phi}(\omega,k)\: \sinh \tau(\omega,k) \right) \\
&=& \frac{i\, \gamma^r}{4 \pi^3\, r}\:
\int_{-\infty}^{\infty}\, d \omega\, e^{i \omega t}
\int_{0}^{\infty} k\, dk\: \left( \cos(kr) - \frac{\sin(kr)}{k r} \right)\; \hat{\phi}(\omega,k)\: \sinh \tau(\omega,k) \:,
\end{eqnarray*}
where we set~$\gamma^r = (\vec{\xi} \vec{\gamma})/|\vec{\xi}|$. Combining
the above terms, we obtain
\begin{eqnarray}
\lefteqn{ P(\xi) \;=\; \frac{1}{4 \pi^3\, r} \int_{-\infty}^{\infty}\, d \omega\, e^{i \omega t}
\int_{0}^{\infty} k\, dk \; \hat{\phi}(\omega, k) \, \Big[ \1 \sin(kr) } \nonumber \\
&& - \,\gamma^0 \, \cosh \tau(\omega, k) \,\sin (kr)\:
+ i \gamma^r\,\sinh \tau(\omega, k)  \,\Big( \!\cos(kr) - \frac{\sin(kr)}{k r} \Big) \Big] . \label{position}
\end{eqnarray}
Note that this formula has a well defined limit as~$r \searrow 0$, and thus we set
\beq \label{origin}
P(t,r=0) \;=\; \frac{1}{4 \pi^3} \int_{-\infty}^{\infty}\, d \omega\, e^{i \omega t}
\int_{0}^{\infty} k^2\, dk \; \hat{\phi}(\omega, k) \, \left[ \1 - \,\gamma^0 \, \cosh \tau(\omega, k) \right] .
\eeq

In~(\ref{momentum}) and~(\ref{position}), the factors~$\gamma^k$
and~$\gamma^r$ involve an angular dependence.
But all the other functions depend only on the position variables~$(t,r)$
and the corresponding momenta~$(\omega, k)$. We now discretize these variables.
In view of~(\ref{tsymm}) it suffices to consider the case~$t \geq 0$.
The position variables should be on a finite lattice~$\Lat$,
\[ (t,r) \;\in\; \Lat \;:=\;
\Big\{0, \Delta_t, \ldots, \,(N_t-1) \Delta_t \Big\} \times
\Big\{ 0, \Delta_r, \ldots, (N_r-1) \,\Delta_r \Big\}\:, \]
where~$N_t$ and~$N_r$ denote the number of lattice points in time and radial directions, and~$\Delta_t, \Delta_r > 0$ are the respective lattice spacings.
The momentum variables should be on the corresponding dual lattice~$\hat{\Lat}$,
\beq \label{omp}
(\omega,k) \;\in\; \hat{\Lat} \;:=\;
\Big\{-(N_t-1) \,\Delta_\omega, \ldots, -\Delta_\omega,0 \Big\} \times
\Big\{ \Delta_k, \ldots, N_r \,\Delta_k \Big\}\:,
\eeq
where we set
\[ \Delta_\omega \;=\; \frac{2 \pi}{\Delta_t\, N_t}\:,\qquad
\Delta_k \;=\; \frac{2 \pi}{\Delta_r\, N_r}\:. \]
We point out that the parameter~$\omega$ in~(\ref{omp}) is
non-positive; this is merely a convention because we are always free to
add to~$\omega$ a multiple of~$N_t \Delta_\omega$.
Furthermore, note that the points with~$k=0$ have been excluded in~$\hat{\Lat}$. This is because the integrands in~(\ref{position}) and~(\ref{origin}) vanish
as~$k \searrow 0$, and thus it seems unnecessary to consider the points with~$k=0$.
However, since~$P(\xi)$ has a non-trivial value at~$r=0$ (see~(\ref{origin})),
it seems preferable to take into account the points with~$r=0$ in the
lattice~$\Lat$.
Replacing the Fourier integrals by a discrete Fourier sum, (\ref{position}) and~(\ref{origin}) become
\begin{eqnarray}
P(\xi) &=& \frac{\Delta_\omega \Delta_k}{4 \pi^3\: r} \sum_{(\omega, k) \in \hat{\Lat}}
 e^{i \omega t}\:k\; \hat{\phi} \;\Big[ (\1 - \gamma^0 \, \cosh \tau)\, \sin(kr) \nonumber \\
&&\spc\qquad\qquad +\, i \gamma^r\,\sinh \tau  \,\Big( \!\cos(kr) - \frac{\sin(kr)}{k r} \Big) \Big],\qquad
{\mbox{if $r \neq 0$}} \label{rnn} \\
P(t, r=0) &=& \frac{\Delta_\omega \Delta_k}{4 \pi^3}
\sum_{(\omega, k) \in \hat{\Lat}}
 e^{i \omega t}\:k^2\; \hat{\phi}
\,(\1 - \gamma^0 \, \cosh \tau) \, , \label{rn}
\end{eqnarray}
with functions~$\hat{\phi}$ and~$\tau$ defined on~$\hat{\Lat}$.

\begin{figure}[t]
\begin{center}
\input{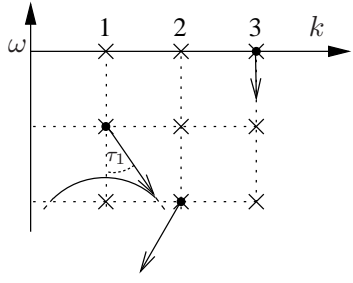}
\caption{Example of a three-particle system on a $3 \times 3$-lattice} \label{fig1}
\end{center}
\end{figure}

The points of the dual lattice~$\hat{\Lat}$ have the interpretation as the
quantum states of the system, which may or may not be occupied by fermionic
particles. More precisely, if~$\hat{\phi}(\omega, p) \neq 0$, a whole ``shell'' of
fermions of energy~$\omega$ and of momenta~$\vec{k}$ with~$|\vec{k}|=p$ is occupied.
For most purposes it is convenient and appropriate to count the whole shell of fermions
as one particle of our lattice model. Thus if~$\hat{\phi}(\omega, p) \neq 0$, we
say that the lattice point~$(\omega, p)$ is {\em{occupied by a particle}};
otherwise the lattice point is not occupied. A system where~$n$ lattice points
are occupied is referred to as an {\em{$n$-particle system}}. Each particle
is characterized by the values of~$\hat{\phi}$ and~$\tau$, or,
equivalently, by the vector $(-2k \hat{\phi} \cosh \tau, 2k \hat{\phi} \sinh \tau)$. It is convenient to
describe the fermion system by drawing these vectors at all occupied lattice
points, as shown in Figure~\ref{fig1} for a three-particle system.

We conclude this section by a discussion of what the parameter~$f$ and the
idempotence condition~$P^2=P$ of discrete space-time mean in the setting of
our lattice model. In discrete space-time, the number of particles~$f$ equals
the trace of~$P$. Computing the trace of~$P$ naively for our lattice model,
our homogeneous ansatz~(\ref{homogeneous}) yields
\beq \label{naiv}
f \;=\; \Tr P \;=\; \int_{\R^4} \tr P(x,x)\, d^4x \;=\; \tr P(\xi=0)\,\cdot\, \infty\:,
\eeq
where~``$\tr$'' denotes the trace of a $4 \times 4$-matrix. According to~(\ref{rn}),
\beq \label{lt}
\tr P(\xi=0) \;=\; \frac{\Delta_\omega \Delta_k}{\pi^3}
\sum_{(\omega, k) \in \hat{\Lat}}
k^2\; \hat{\phi}(\omega, k) \:,
\eeq
showing that~(\ref{naiv}) is equal to~$+\infty$ unless~$P$ vanishes identically.
Here we used essentially that, although~$\xi=y-x$ was discretized
on a finite lattice, the space-time variable~$x$ itself is still an arbitrary
vector in Minkowski space. In other words, our lattice system is a homogeneous
system in infinite volume, and in such a system the number of particles is
necessarily infinite.
The simplest way to bypass this problem is to note that for a
homogeneous system in discrete space-time~\cite[Def.~2.4]{F1},
\[ f \;=\; \sum_{x \in M} \Tr (E_x P) \;=\; m\, \Tr(E_1 P)\:, \]
and so the number of particles grows linearly with the number of
space-time points. Due to this simple connection, we can disregard~$f$
and consider instead the local trace. This has the advantage that
the local trace can be identified with the expression~(\ref{lt}) of our
lattice system. For the variational principle in discrete space-time~(\ref{Ldef},
\ref{Sdef}), it is important that variations of~$P$ keep the number of
particles~$f$ fixed. This condition can be carried over to our lattice
system, giving rise to the so-called {\em{trace condition}} (TC):
\begin{description}
\item[(TC)] When varying the fermionic projector of the lattice system~(\ref{rnn},
\ref{rn}), the local trace as defined by
\[ \floc \;:=\;
\frac{\Delta_\omega \Delta_k}{\pi^3}
\sum_{(\omega, k) \in \hat{\Lat}}
k^2\; \hat{\phi}(\omega, k) \]
should be kept fixed.
\end{description}
We conclude that, although~$f$ is infinite for our lattice system, the
local trace~$\floc$ is well defined and finite. This all we need, because
with~(TC) we have implemented the condition corresponding to the
condition in discrete space-time that~$f$ should be
kept fixed under variations of~$P$. We point out that neither~$f$
nor~$\floc$ coincides with the number of particles
as obtained by counting the occupied states.

The idempotence condition~$P^2=P$ is satisfied if and only if the
fermionic wave functions are properly normalized.
As explained above, our lattice model is defined in
infinite space-time volume, and thus a-priori the normalization integrals diverge.
As shown in~\cite[\S2.6]{PFP}, a possible method for removing this
divergence is to consider the system in finite $3$-volume and to smear out the
mass parameter. However, there are other normalization methods, and it is not
clear whether they all give rise to the same normalization condition for our lattice
model. The basic difficulty is related to the fact that each occupied lattice
point~$(\omega, p) \in \hat{\Lat}$ corresponds to a whole shell of fermions
(see above). Thus the corresponding summand in~(\ref{rnn}, \ref{rn}) involves
an ``effective wave function'' describing an ensemble of fermions. But it is
not clear of how many fermions the ensemble consists and thus, even if we knew
how to normalize each individual fermion, the normalization of the effective
wave function would still be undetermined. This problem becomes clear if one
tries to model the same physical system by two lattice models with two different
lattice spacings. Then in general one must combine several occupied lattice
points of the finer lattice to one ``effective'' occupied lattice point of the
coarser lattice. As a consequence, the normalization of the coarser lattice must
be different from that on the finer lattice. This explains why there is no simple
canonical way to normalize the effective wave functions.

\begin{figure}[t]
\begin{center}
\input{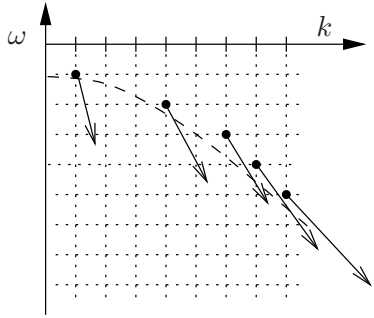}
\caption{A discretized Dirac sea} \label{fig2}
\end{center}
\end{figure}

Our method for avoiding this normalization problem is to choose the normalization
in such a way that the fermionic projector of the continuum can be carried over easily
to the lattice system:
In Minkowski space, a Dirac sea in the vacuum is described by the
distribution (see~\cite[\S2.2]{PFP})
\beq \label{Pcont}
\hat{P}(p) \;=\; (p \slsh+m)\, \delta(p^2-m^2)\: \Theta(-p^0)\:.
\eeq
Taking the Fourier transform and carrying out the angular integrals,
we obtain again the expressions~(\ref{position}, \ref{origin}), but now
with~$\hat{\phi}(\omega, k) = \delta(\omega^2-k^2-m^2)$.
This allows us to carry out the~$k$-integral,
\[ \int_{-\infty}^{\infty}\, d \omega\, e^{i \omega t}
\int_{0}^{\infty} k\, dk \; \delta(\omega^2-k^2-m^2) \;\cdots \;=\;
\int_{\R \setminus [-m,m]}\, d \omega\, e^{i \omega t}\;\frac{1}{2}\:
\cdots \Big|_{k=\sqrt{\omega^2-m^2}} \:. \]
The easiest method to discretize the obtained expression is to
replace the $\omega$-integral by a sum, and to choose
for every $\omega \leq -m$ a lattice point~$(\omega, k) \in \hat{\Lat}$
such that
\beq \label{dDs}
0 \;\leq\;  k - \sqrt{\omega^2-m^2} \;<\; \Delta_k \:.
\eeq
An example for the resulting discretized Dirac sea
is shown in Figure~\ref{fig2}.

Note that for this configuration, $\hat{\phi}(\omega,k)=1/(2k)$
at all occupied lattice points.
Next we allow to modify this configuration, as long as the
normalization integrals remain unchanged:
First, changing~$\tau$ corresponds to a unitary transformation
of the corresponding state, without influence on the normalization.
Second, hopping from a lattice point to another unoccupied lattice point
with the same value of~$k$ changes the state only by the phase factor~$\exp(-i
(\omega-\omega') t)$, again without influence on the normalization.
This leaves us with the so-called {\em{normalization condition}} (NC):
\begin{description}
\item[(NC)] The function~$\hat{\phi}$ in~(\ref{rnn}, \ref{rn})
should only take the two values
\[ \hat{\phi}(\omega,k) \;=\; 0 \spc {\mbox{or}} \spc
\hat{\phi}(\omega,k) \;=\; \frac{1}{2 k}\:. \]
\end{description}
We again point out that this normalization condition is not canonical.
It could be modified or even be left out completely. It seems an interesting
question to analyze how the behavior of the lattice model depends on the
choice of the normalization condition.

\section{The Variational Principle on the Lattice} \label{sec4}
The Lagrangian~(\ref{Ldef}) is also well defined for our lattice model.
Let us compute it in more detail. We decompose the fermionic
projector~(\ref{position}, \ref{origin}) into its scalar and vector
components,
\[ P(\xi) \;=\; \phi(t,r)\: \1 + v_0(t,r) \,\gamma^0 + v_r(t,r) \,\gamma^r
\;=\; \phi \, \1 + v_j\, \gamma^j \:. \]
Furthermore, using that the functions~$\hat{\phi}$ and~$\tau$ are real,
we find that
\[ P(-\xi) \;=\; P(\xi)^* \;=\; \overline{\phi} \, \1 + \overline{v_j}\, \gamma^j \:. \]
Thus, omitting the argument~$\xi$, the closed chain~(\ref{cdef})
becomes
\[ A \;=\; (\vslsh + \phi)(\overline{\vslsh} + \overline{\phi})\:. \]
For the computation of the spectrum, it is useful to decompose~$A$ in the form
\[ A \;=\; A_2 + A_2 + \mu \]
with
\[ A_1 \;=\; \frac{1}{2} \left[ \vslsh, \overline{\vslsh} \right]\:,\quad
A_2 \;=\; \phi\, \overline{\vslsh} + \vslsh \,\overline{\phi}\:,\quad
\mu \;=\; v_j \overline{v^j} + \phi \overline{\phi}\:. \]
A short calculation shows that the matrices~$A_1$ and~$A_2$ anti-commute, and thus
\beq \label{Arel}
(A-\mu)^2 \;=\; A_1^2 + A_2^2 \;=\; D[A]\: \1\:,
\eeq
where we set
\beq \label{discriminant}
D[A] \;=\; \frac{1}{4}\: \tr (A^2) - \frac{1}{16}\: (\tr A)^2
\;=\; (v_j \,\overline{v^j})^2 - |v_j \,v^j|^2
+ (v_j \,\overline{\phi} + \phi \,\overline{v_j})
\:(v^j \,\overline{\phi} + \phi \,\overline{v^j})\:.
\eeq
The identity~(\ref{Arel}) shows that the characteristic polynomial of the matrix~$A$ has
the two zeros
\beq \label{lambdas}
\lambda_\pm \;=\; v_j \,\overline{v^j} + \phi \overline{\phi} \pm \sqrt{ D } \:.
\eeq
If these two zeros are distinct, they both have multiplicity two. If the two zeros coincide, there
is only one zero of multiplicity four. Hence the Lagrangian~(\ref{Ldef}) simplifies to
\beq \label{Lags}
{\mathcal{L}}[A] \;=\; \left( |\lambda_+| - |\lambda_-| \right)^2
\eeq
In order to further simplify the Lagrangian, we introduce
a discrete causal structure, in agreement with~\cite{F5}.
\begin{Def} \label{defcausal}
A lattice point~$(t,r) \in \Lat$ is called
\[ \left. \begin{array}{ccl} {\mbox{timelike}} &\;\;\;& {\mbox{if $D[A(t,r)] \;\geq\; 0$}} \\
{\mbox{spacelike}} && {\mbox{if $D[A(t,r)] \;<\; 0\:.$}}
\end{array} \right. \]
\end{Def}
If~$(t,r)$ is spacelike,
the~$\lambda_\pm$ form a complex conjugate pair, and the Lagrangian~(\ref{Lags})
vanishes. If conversely the discriminant is non-negative,
the~$\lambda_\pm$ are both real. In this case, the calculation
\begin{eqnarray*}
\lambda_+ \lambda_- &=& (v \overline{v} + \phi \overline{\phi})^2 - \left[ (v \overline{v})^2
- v^2\: \overline{v}^2 + (v \overline{\phi} + \phi \overline{v})^2 \right] \\
&=& 2\:(v \overline{v})\: |\phi|^2 + |\phi|^4 + v^2\: \overline{v}^2 -
(v \overline{\phi} + \phi \overline{v})^2 \\
&=& |\phi|^4 + v^2\: \overline{v}^2 - v^2\: \overline{\phi}^2 - \phi^2\: \overline{v}^2
\;=\; (v^2-\phi^2)(\overline{v}^2 - \overline{\phi}^2) \;\geq\; 0
\end{eqnarray*}
(where we omitted the tensor indices in an obvious way)
shows that~$\lambda_+$ and~$\lambda_-$ have the same sign, and so we can leave out the absolute values in~(\ref{Lags}). We conclude that
\[ {\mathcal{L}}(t,r) \;=\; \left\{
\begin{array}{cl} 4 D[A(t,r)] & {\mbox{if~$(t,r)$ is timelike}} \\
0 & {\mbox{otherwise ,}} \end{array} \right. \]
where~$D$ is given by~(\ref{discriminant}).
Hence our Lagrangian is compatible with the discrete causal structure
in the sense that it vanishes if~$(t,r)$ is spacelike.

Before we can set up the variational principle, we need to think about
what the sum over the space-time points in~(\ref{Sdef}) should
correspond to in our lattice system. Since we are considering a homogeneous
system, one of the sums simply gives a factor~$m$, and we can leave out this
sum. The other sum in the continuum
should correspond to a space-time integral (see for example~(\ref{ip})).
In our lattice system, the point~$(t,r)$ can be thought of as the
$2$-dimensional sphere~$|\vec{\xi}|=r$ at time~$t$. Therefore,
we replace the spatial integral by a sum over the
discretized radii, but with a weight factor which takes into account
that the surfaces of the spheres grow quadratically in~$r$.
More precisely, we identify~$(t,r)$ with a shell
of radius between~$r-\Delta_r/2$ and~$r+\Delta_r/2$.
This leads us to the replacement rule
\[ \int_{\R^3} d\vec{\xi}\: \cdots \;\longrightarrow\;
\Delta_r^3 \sum_{n=0}^{N_r-1} \rho_r(n \Delta_r) \:\cdots \]
with the weight function~$\rho_r$ given by
\beq \label{weight}
\rho_r(n \Delta_r) \;=\; \frac{4 \pi}{3} \cdot \left\{
\begin{array}{cl} \displaystyle 1/8 & {\mbox{if~$n=0$}} \\[.3em]
\displaystyle \left(n + 1/2 \right)^3 - \left(n - 1/2 \right)^3
& {\mbox{if~$n>0\:.$}} \end{array} \right.
\eeq
When discretizing the time integral, we need to take into account
that on the lattice~$\Lat$, the time parameter~$t$ is always
non-negative. Since the Lagrangian is symmetric,
${\mathcal{L}}[A_{xy}] = {\mathcal{L}}[A_{yx}]$
(see~\cite[\S3.5]{PFP}), this can be done simply by counting
the lattice points with~$t>0$ twice. Thus we discretize the time
integral by
\[ \int_{-\infty}^\infty dt\: \cdots \;\longrightarrow\;
\Delta_t \sum_{n=0}^{N_r-1} \rho_t(n \Delta_r) \:\cdots \]
with
\beq \label{tweight}
\rho_t(n \Delta_t) \;=\; \left\{
\begin{array}{cl} 1 & {\mbox{if~$n=0$}} \\
2 & {\mbox{if~$n>0\:.$}} \end{array} \right.
\eeq
Then the action becomes
\[ {\mathcal{S}}[P] \;=\; \Delta_t \:\Delta_r^3
\sum_{(t,r) \in \Lat} \rho_t(t)\, \rho_r(r)\: {\mathcal{L}}(t,r)\:. \]
Our variational principle is to minimize this action
by varying the functions~$\hat{\phi}$ and~$\tau$ in~(\ref{rnn}, \ref{rn})
under the constraints~(TC) and~(NC).

With the constructions of Sections~\ref{sec3} and~\ref{sec4} we successively
derived our two-dimensional lattice model. Clearly, not all the arguments
leading to the model were rigorous, and also we put in strong assumptions
on the physical situation which we have in mind.  More precisely, the main
assumption was the spherically symmetric and static ansatz with a
vector-scalar structure~(\ref{fourier}, \ref{vss}); this ansatz was
merely a matter of convenience and simplicity. Moreover, the choice of
the weight function~$\rho$ involved some arbitrariness. However, we do not
consider this to be critical because choosing the weight factors
in~(\ref{weight}) differently should not change the qualitative behavior
of the model (except that for the existence of minimizers it is important
that~$\rho_r(0) \neq 0$; see Section~\ref{sec6}).
Finally, the normalization condition~(NC) could be modified, as discussed
in detail at the end of Section~\ref{sec3}.

The main point of interest of our lattice model is that it allows
to describe discretizations of Dirac seas~(\ref{dDs}) but also completely
different configurations of the fermions. Thus within the lattice model it
should be possible to analyze in detail whether and how Dirac sea
configurations form as minimizers of our variational principle.
Moreover, in our lattice model one can implement all the spherically
symmetric regularization effects as found in~\cite{F4}. Hence our lattice
model should make it possible to verify effects from~\cite{F4} coming from
the discrete side and to analyze these effects in greater detail.

In the next section we shall define our lattice model once again more
systematically, making the following simplifications:
\begin{itemize}
\item By scaling we can always arrange that~$\Delta_\omega$ and~$\Delta_k$ have
an arbitrary value. It is most convenient to choose
\[ \Delta_\omega \;=\; 1\:,\qquad \Delta_k \;=\; 1 \:. \]
Then
\[ \Delta_t \;=\; \frac{2 \pi}{N_t}\:,\qquad
\Delta_r \;=\; \frac{2 \pi}{N_r}\:. \]
\item The formulas for~$P$, (\ref{rnn}, \ref{rn}), only involve the two
Dirac matrices~$\gamma^0$ and~$\gamma^r$, which satisfy the anti-commutation
rules
\[ (\gamma^0)^2 \;=\; \1\:,\quad (\gamma^r)^2 \;=\; -\1\:,\quad
\left\{ \gamma^0, \gamma^r \right\} \;=\; 0 \:. \]
Since these anti-commutation rules can be realized already by $2 \times 2$-matrices,
we may simplify the matrix structure by the replacements
\[ \gamma^0 \;\longrightarrow\; \sigma^3 \:,\qquad
\gamma^r \;\longrightarrow\; -i \sigma^1 , \]
where~$\sigma^i$ are the usual Pauli matrices. Modifying the definition of
the discriminant~(\ref{discriminant}) to
\[ D[A] \;=\; \frac{1}{2}\: \tr (A^2) - \frac{1}{4}\: (\tr A)^2 \]
(where now~``$\tr$'' clearly denotes the trace of a $2 \times 2$-matrix), the
Lagrangian remains unchanged.
\item In order to simplify the normalization condition~(NC), it is convenient
to introduce the function
\[ \Phi(k, \omega) \;=\; 2 k\, \hat{\phi}(k,\omega)\:, \]
where for notational simplicity we also omitted the tilde.
\item In order to simplify the prefactors, we multiply~$P$ by~$4 \pi$,
divide the Lagrangian by four, and divide the action by~$(2 \pi)^4\, \pi/6$.
Furthermore, we multiply~$\floc$ by a factor of~$\pi^3$.
\end{itemize}

\section{Definition of the Model and Basic Properties} \label{sec5}
For given integer parameters~$N_t$, $N_r$ and~$p$ we introduce
the lattice~$\Lat$ and its dual lattice~$\hat{\Lat}$,
\begin{eqnarray*}
(t,r) \;\in\; \Lat &=&
\Big\{0,\, \frac{2 \pi}{N_t},\, \ldots, \, 2 \pi\, \frac{N_t-1}{N_t} \Big\} \times
\Big\{ 0,\, \frac{2 \pi}{N_r}, \ldots, 2 \pi\, \frac{N_r-1}{N_r} \Big\} \\
(\omega,k) \;\in\; \hat{\Lat} &=&
\Big\{-(N_t-1),\, \ldots, \,-1,\, 0 \Big\} \times
\Big\{ 1,\, \ldots,\, N_r \Big\}\:.
\end{eqnarray*}
On~$\hat{\Lat}$ we choose a non-negative function~$\Phi$ and a real function~$\tau$, which vanish
except at~$p$ lattice points. We set
\begin{eqnarray}
P(\xi) &=& \frac{1}{r} \sum_{(\omega, k) \in \hat{\Lat}}
 e^{i \omega t}\:\Phi \;\Big[ (\1 - \sigma^3 \, \cosh \tau)\, \sin(kr) \nonumber \\
&&\spc\qquad\qquad +\, \sigma^1\,\sinh \tau  \,\Big( \!\cos(kr) - \frac{\sin(kr)}{k r} \Big) \Big],\qquad
{\mbox{if $r \neq 0$}} \label{Pdef} \\
P(t, r=0) &=& \sum_{(\omega, k) \in \hat{\Lat}}
 e^{i \omega t}\:k\; \Phi
\,(\1 - \sigma^3 \, \cosh \tau) \,, \label{Podef}
\end{eqnarray}
where~$\sigma^1$ and~$\sigma^3$ are two Pauli matrices.
For any~$(t,r) \in \Lat$ we introduce the closed chain~$A(t,r)$ by
\[ A(t,r) \;=\; P(t,r)\,P(t,r)^* \:, \]
where the adjoint with respect to the spin scalar product is
given by
\[ P(t,r)^* \;=\; \sigma^3\, P(t,r)^\dagger\, \sigma^3 \:, \]
and the dagger denotes transposition and complex conjugation.
We define the discriminant~$D[A]$ and the Lagrangian~${\mathcal{L}}[A]$ by
\begin{eqnarray}
D[A] &=& \frac{1}{2}\: \tr (A^2) - \frac{1}{4}\: (\tr A)^2 \label{Ddef} \\
{\mathcal{L}}[A] &=& D[A]\: \Theta(D[A])\:, \label{Lagdef}
\end{eqnarray}
where~$\Theta$ is the Heaviside function. The action is
\[ {\mathcal{S}} \;=\; \frac{1}{N_t N_r^3}
\sum_{(t,r) \in \Lat} \rho_t(t)\, \rho_r(r)\, {\mathcal{L}}[A(t,r)] \:, \]
where~$\rho_t$ and~$\rho_r$ are the weight functions
\begin{eqnarray*}
\rho_t\!\left(2 \pi\, \frac{n}{N_t} \right) &=& \left\{
\begin{array}{cl} 1 & {\mbox{if~$n=0$}} \\[.3em]
2 & {\mbox{if~$n>0$}} \end{array} \right. \\
\rho_r\!\left(2 \pi\, \frac{n}{N_r} \right) &=& \left\{
\begin{array}{cl} 1 & {\mbox{if~$n=0$}} \\[.3em]
\left(2n + 1 \right)^3 - \left(2n - 1 \right)^3 &
{\mbox{if~$n>0\:.$}} \end{array} \right.
\end{eqnarray*}
Our variational principle is to minimize the action, varying the functions~$\Phi$
and~$\tau$ under the following constraints:
\begin{description}
\item[(TC)] The local trace
\beq \label{ltracedef}
\floc \;:=\; \sum_{(\omega, k) \in \hat{\Lat}} k\; \Phi(\omega, k)
\eeq
should be kept fixed.
\item[(NC)] The function~$\Phi$
should only take the two values~$\Phi(\omega,k) = 0$ or~$\Phi(\omega,k) = 1$.
\end{description}
The last condition~(NC) could be weakened or left out
(see the discussion at the end of Section~\ref{sec3}).

According to Definition~\ref{defcausal}, the functions~$\Phi$ and~$\tau$
induce on~$\Lat$ a {\em{discrete causal structure}}.
The Lagrangian is compatible with the discrete causal structure
in the sense that it vanishes if~$(t,r)$ is spacelike.
Furthermore, our lattice system has the following symmetries:
\begin{description}
\item[symmetry under parity transformations:] The traces in~(\ref{Ddef}) vanish unless an even number
of matrices~$\sigma^1$ appears. Therefore, the Lagrangian remains unchanged
if the factor~$\sinh \tau$ in~(\ref{Pdef}) flips sign. Hence the action
is symmetric under the transformation
\beq \label{parity}
\tau(\omega, k) \;\longrightarrow\; -\tau(\omega, k) \qquad
{\mbox{for all $(\omega, k) \in \hat{\Lat}$}}.
\eeq
This transformation changes the sign of the spatial component of~$P$.
The name ``parity transformation'' comes from the analogy to the
usual parity transformation~$\vec{x} \rightarrow -\vec{x}$.
\item[gauge symmetry:] We introduce on the dual lattice~$\hat{\Lat}$
for any~$\Omega \in \Z$ the translation respecting the periodic boundary conditions
\beq \label{gauge}
\omega \;\longrightarrow\; \tilde{\omega} \;=\; \left( \omega + \Omega \right) {\mbox{ mod }} N_t
\eeq
and also translate the functions~$\tau$ and~$\Phi$ by setting
\[ \tilde{\tau}(\tilde{\omega},k) \;=\; \tau(\omega, k)\:,\quad
\tilde{\Phi}(\tilde{\omega},k) \;=\; \Phi(\omega, k)\:. \]
This translation in momentum space corresponds to a multiplication
by a phase factor in position space,
\[ \tilde{P}(\xi) \;=\; e^{i \Omega t}\, P(\xi)\:. \]
This phase factor drops out when forming the closed chain, and thus the
Lagrangian remains unchanged.
The transformation~(\ref{gauge}) are precisely those local gauge
transformations which are compatible with our spherically symmetric and static ansatz.
\end{description}

\section{Existence of Minimizers} \label{sec6}
In this section we prove an existence result, which is so general that it
applies also in the case when the normalization condition~(NC) is weakened.

\begin{Prp} Consider the variational principle of Section~\ref{sec5}
with the trace condition~(TC) and, instead of~(NC), the weaker
condition that that there is a parameter~$\varepsilon>0$ such that
\[ \Phi(\omega,k) \;=\; 0 \quad {\mbox{or}} \quad
\Phi(\omega, k) \;>\; \varepsilon \spc
{\mbox{for all~$(\omega, k) \in \hat{\Lat}$}}\,. \]
Then the minimum of the action is attained.
\end{Prp}
{\Proof} Since the Lagrangian is non-negative, we can estimate the action
from above by the Lagrangian at the origin~$t=r=0$,
\beq \label{Ses}
{\mathcal{S}} \;\geq\; {\mathcal{L}}[A(0,0)]\:.
\eeq
At the origin, the fermionic projector takes the form (see~(\ref{Podef}))
\[ P(0) \;=\; \sum_{(\omega, k) \in \hat{\Lat}}
k\; \Phi \,(\1 - \sigma^3 \, \cosh \tau) \:. \]
This matrix can be diagonalized and has the two eigenvalues
\[ \mu_\pm \;=\; \sum_{(\omega, k) \in \hat{\Lat}}
k\; \Phi \,(1 \pm \cosh \tau) \:. \]
Thus the closed chain~$A(0,0)$ has the two
eigenvalues~$\lambda_\pm = \mu_\pm^2$. As a consequence,
using~(\ref{Ddef}) and~(\ref{ltracedef}),
\begin{eqnarray}
{\mathcal{L}}[A(0,0)] &=&
\frac{1}{4} \left(\lambda_+ - \lambda_- \right)^2 \;=\;
\frac{1}{4} \left(\mu_+ + \mu_- \right)^2 \left(\mu_+ - \mu_- \right)^2 \nonumber \\
&=& 4\,\floc^2 \left( \sum_{(\omega, k) \in \hat{\Lat}}
k\; \Phi \, \cosh \tau \right)^2 . \label{Les}
\end{eqnarray}

Consider a minimal sequence. Then, according to~(\ref{Ses}),
the expression~(\ref{Les}) is uniformly bounded. If~$\floc=0$,
our system is trivial, and thus we may assume that~$\floc$ is
a positive constant. Using~(\ref{ltracedef}) and the fact
that~$k \geq 1$, we conclude that the functions~$\Phi$ are uniformly
bounded. The boundedness of~(\ref{Les}) implies that
there is a constant~$C>0$ such that
\[ \sum_{(\omega, k) \in \hat{\Lat}}
k\; \Phi \, \cosh \tau \;\leq\; C \]
for all elements of the minimal sequence.
Whenever~$\Phi$ vanishes, we can also set~$\tau$ equal to zero.
If~$\Phi$ is non-zero, the inequality~$\Phi \geq \varepsilon$
gives a uniform upper bound for~$\cosh \tau$,
\[ \cosh \tau \;\leq\; \frac{C}{\varepsilon}\:. \]
We conclude that the functions~$\Phi$ and~$\tau$ are uniformly bounded.
Hence a compactness argument allows us to choose a convergent subsequence.
Since our action is obviously continuous, the limit is the desired
minimizer.
\QED
We point out that this proposition makes no statement on uniqueness.
There seems no reason why the minimizers should be unique. In Section~\ref{sec7}
we shall see examples with several minimizers.

\section{First Numerical Results, A Mechanism of Spontaneous Symmetry Breaking} \label{sec7}
\begin{figure}[t]
\begin{center}
 \includegraphics[width=10cm]{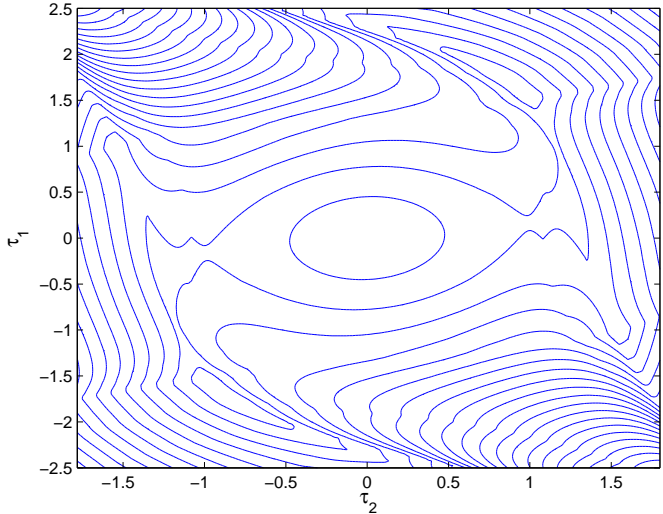}
 \caption{Action for a two-particle system on an $8 \times 6$-lattice} \label{fig3}
\end{center}
\end{figure}

We now discuss numerical results for
a simple lattice system, with the intention of exemplifying a few
general properties of our model. More precisely, we
choose~$N_t=8$, $N_r=6$ and occupy two lattice points, one with momentum~$k=1$ and
one with~$k=2$. Satisfying the condition~(TC) and~(NC), the system is
characterized by the two discrete parameters~$\omega_1, \omega_2 \in
\{-7,\ldots,0\}$ and the two real parameters~$\tau_1$ and~$\tau_2$ at the occupied lattice
points. We find that the absolute minimum of the action is attained 
if we choose~$\omega_1=-1$ and~$\omega_2=-2$. For this choice of the
occupied lattice points, Figure~\ref{fig3} shows a contour plot of the action in
the~$(\tau_1, \tau_2)$-plane.
The plot is symmetric under reflections at the origin; this is the symmetry under parity
transformations~(\ref{parity}). The minimum at the origin corresponds to the trivial
configuration~$\tau_1 = \tau_2 = 0$, where the two vectors~$v_i :=(-\Phi_i \cosh \tau_i,
\Phi_i \sinh \tau_i)$ are both parallel to the $\omega$-axis.
However, this is only a local minimum, whereas
the absolute minimum of the action is attained at the two
points~$(\tau_1 \approx 1.5, \tau_2 \approx 1)$
and~$(\tau_1 \approx -1.5, \tau_2 \approx -1)$.
The interesting point is that the absolute minima are non-trivial,
meaning that the occupied points distinguish specific lattice points and that
the corresponding vectors~$v_i$ are not parallel. The minimizers are not symmetric
under~(\ref{parity}). Thus by choosing one of the minimizers, the symmetry under parity
transformations is spontaneously broken.

This effect of spontaneous symmetry breaking can be understood in analogy to the Higgs
mechanism in the standard model, where the double-well potential has non-trivial minima,
and by choosing such a minimizer the corresponding vacuum breaks the original
$SU(2)$-symmetry. For our lattice system, the role of the Higgs potential is played
by the particular form of our action in combination with the constraints as given by the
trace condition and the normalization condition.

We remark for clarity that the effect of spontaneous symmetry breaking observed here is
much different from the effect described in~\cite{F3} for general fermion systems in
discrete space-time. Apart from the fact that in~\cite{F3} we consider instead of parity
the outer symmetry group, the main difference is that in~\cite{F3} we do not specify the
action, but the spontaneous symmetry breaking arises merely by constructing fermionic projectors
for a given outer symmetry group. Here, on the contrary, the specific form of the action is
crucial. This is made clear by the alternative action ${\mathcal{S}}=\sum_{(\omega, k)
\in \hat{\Lat}} |\tau(\omega, k)|$, which only has trivial minimizers with~$\tau \equiv 0$
which are all symmetric under the parity transformation.

We conjecture that for a large lattice and many particles there are minimizers which
look similar to the discretized Dirac sea structure of Figure~\ref{fig2}. Since the
presentation of larger simulations is more involved and also requires a detailed description of the
numerical methods, we postpone this analysis to a forthcoming publication.

\noindent
NWF I -- Mathematik,
Universit{\"a}t Regensburg, 93040 Regensburg, Germany, \\
{\tt{Felix.Finster@mathematik.uni-regensburg.de}}, \\
{\tt{Waetzold.Plaum@mathematik.uni-regensburg.de}}

\end{document}